\documentclass[aps, nofootinbib,superscriptaddress, preprintnumbers, longbibliography]{article}

\usepackage{tikz}
\usepackage{jcappub}
\usepackage[utf8]{inputenc}
\usepackage{amsmath,amsfonts,amssymb}
\usepackage{graphicx}
\usepackage[toc,page]{appendix}
\usepackage{cleveref}
\usepackage[loose]{units}
\usepackage{booktabs}
\usepackage{mathtools}
\usepackage{listings}
\usepackage{xcolor}

\makeatletter
\gdef\@fpheader{}
\makeatother
\begin{document}
\title{
	\texttt{InflationEasy}:\\ A C++ Lattice Code for Inflation

    \vspace{0.5cm}

\begin{center} 
    \centering
    \includegraphics[width=1\textwidth]{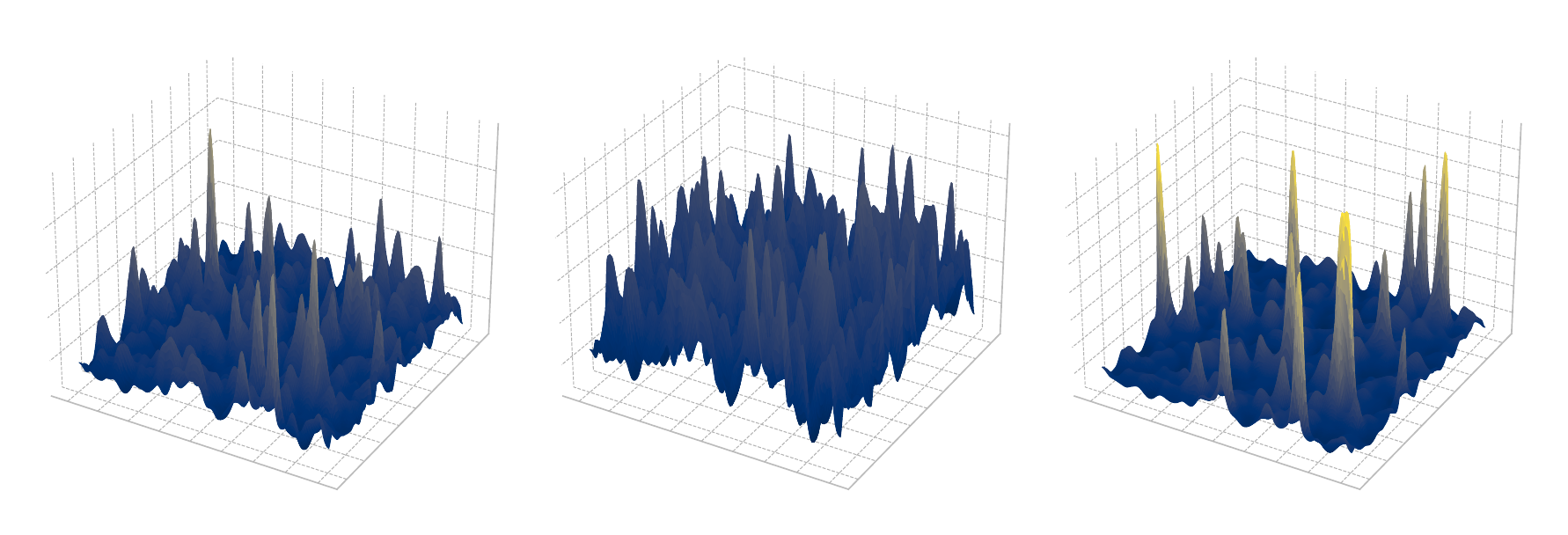}
\end{center}

 \vspace{-2cm}
}
\newcommand{\grappa}{Gravitation Astroparticle Physics Amsterdam (GRAPPA),\\University of Amsterdam, Science Park 904, 1098 XH Amsterdam, The Netherlands}

\author{Angelo Caravano}
\affiliation{\grappa}
\emailAdd{a.caravano@uva.nl}

\abstract{
\href{https://github.com/caravangelo/inflation-easy.git}{\texttt{InflationEasy}} is a lattice code specifically developed for cosmological inflation. It simulates the nonlinear dynamics of a scalar field on a three-dimensional lattice in an expanding FLRW universe using finite-difference spatial derivatives. Based in part on the well-known \texttt{LATTICEEASY}, it incorporates several features tailored specifically to inflationary applications, including a nonperturbative $\delta N$ method to compute the curvature perturbation at the end of inflation $\zeta$ directly from the lattice. In addition to the scalar sector, the code can also simulate scalar-induced gravitational waves, accounting for contributions generated both during inflation and during the subsequent horizon re-entry of scalar perturbations, and enabling the computation of the resulting gravitational-wave background. \texttt{InflationEasy} enables fully nonlinear studies of regimes with large fluctuations or nonperturbative non-Gaussianities, which lie beyond the reach of standard perturbation theory. It is applicable to a broad range of inflationary models, including those relevant for primordial black hole formation, gravitational-wave backgrounds, and large-scale structure.

}

\maketitle
	
\section{Introduction}

Lattice simulations have become essential tools in studying the nonlinear dynamics of the early universe, particularly in regimes where perturbation theory breaks down, such as preheating and cosmological phase transitions. Over the past two decades, several public codes have been developed for this purpose, including \texttt{LATTICEEASY}~\cite{latticeeasy}, \texttt{DEFROST}~\cite{Frolov_2008}, \texttt{PSpectRe}~\cite{Easther_2010}, \texttt{HLattice}~\cite{hlattice}, \texttt{PyCOOL}~\cite{Sainio_2012}, \texttt{GFiRe}~\cite{Lozanov_2020}, and \texttt{CosmoLattice}~\cite{figueroa2021cosmolattice}. These codes have been widely used to simulate nonlinear field dynamics after inflation, especially during reheating. However, none are specifically designed to simulate the inflationary phase itself, far from the end of inflation.\footnote{An exception is \texttt{STOLAS}~\cite{Mizuguchi:2024kbl}, which simulates inflation using the stochastic inflation formalism, which differs substantially from the standard lattice method of solving differential equations on a fixed comoving grid.}

\texttt{InflationEasy} fills this gap. It is a simple, publicly available C++ code that evolves a canonical scalar field on an expanding three-dimensional lattice during inflation. Designed specifically for inflationary dynamics, it provides an accessible platform to simulate a wide range of single-field models. The code is lightweight, modular, and easy to modify—even for users with minimal experience in C++—and, for modest lattice resolutions (e.g.\ $128^3$--$256^3$), simulations can be run efficiently on an ordinary laptop. Among its key features is the implementation of a $\delta N$ method to compute the comoving curvature perturbation $\zeta$ directly from the lattice. The lattice-based framework of \texttt{InflationEasy} enables fully nonperturbative calculations of inflationary predictions in regimes that lie beyond the regime of applicability of standard perturbative methods.

In addition to scalar observables, \texttt{InflationEasy} can also compute scalar-induced gravitational waves. The code accounts for contributions generated both during inflation and during the subsequent horizon re-entry of scalar perturbations, allowing for fully nonlinear predictions of the resulting stochastic gravitational-wave background. These capabilities make \texttt{InflationEasy} suitable for studying a broad range of inflationary scenarios, including those relevant for primordial black hole formation, gravitational-wave production, and the origin of large-scale structure.

\texttt{InflationEasy} has already been employed in several studies~\cite{Caravano:2021pgc,Caravano:2022yyv,Caravano:2024tlp,Caravano:2024moy,Caravano:2025diq,gwpaper}, where it has been used to explore a variety of inflationary models and to benchmark fully nonlinear calculations of $\zeta$ based on the $\delta N$ formalism. This paper accompanies the public release of the code and serves both as a feature summary and as a user manual. The public version of the code can be downloaded at \href{https://github.com/caravangelo/inflation-easy.git}{https://github.com/caravangelo/inflation-easy.git}. It includes, together with the source code, an example notebook illustrating how to process and visualize the simulation outputs.

The paper is organized as follows. In \cref{sec:eoms}, we present the physical system and its equations of motion. In \cref{sec:implementation}, we describe the numerical implementation, including discretization, time evolution, and output generation. \Cref{sec:instructions} contains user instructions, including a walkthrough of the default options and how to configure a new potential. In \cref{sec:GW}, we describe the computation of scalar-induced gravitational waves, including contributions generated during inflation and during the subsequent horizon re-entry of scalar perturbations. We conclude in \cref{sec:conclusions}.

\section{Equations of motion in continuous space}
\label{sec:eoms}
The code solves the Klein-Gordon equation for a canonical scalar field $\phi$ in a flat FLRW spacetime using conformal time:
\begin{equation}
\label{eq:fullcont}
\partial_\tau^2\phi + 2\mathcal{H}\partial_\tau \phi - \nabla^2\phi + a^2\frac{\partial V}{\partial \phi} = 0,
\end{equation}
where $a(\tau)$ is the scale factor and $\mathcal{H} = \partial_\tau a / a$ is the conformal Hubble parameter. A key feature of the lattice approach is that no background/perturbation split is introduced—the inflaton field and its fluctuations are evolved altogether.
The evolution of the scale factor is governed by the Friedmann equations:
\begin{align}
\label{eq:friedmann}
&\mathcal{H}^2 = \frac{1}{3} \langle \rho \rangle a^2,\\
&\frac{d^2a}{d\tau^2} = \frac{1}{6} \left(\langle \rho \rangle - 3 \langle p \rangle\right) a^3,
\end{align}
where $\langle \rho \rangle$ and $\langle p \rangle$ are the average energy density and pressure:
\begin{equation}
\begin{split}
\rho = -T^0_{\hspace{1mm}0} = \frac{(\partial_\tau\phi)^2}{2a^2} + \frac{|\vec\nabla\phi|^2}{2a^2} + V(\phi),\\
p = \frac{1}{3} \sum_i T^i_{\hspace{1mm}i} = \frac{(\partial_\tau\phi)^2}{2a^2} - \frac{|\vec\nabla\phi|^2}{6a^2} - V(\phi).
\end{split}
\label{eq:ED}
\end{equation}
Throughout this paper we work in reduced Planck units, where the reduced Planck mass
$M_{\rm Pl} \equiv (8\pi G)^{-1/2}$ is set to unity.

In the code, the gradient term is evaluated via the identity
\[
|\vec\nabla\phi|^2 \equiv -\phi \nabla^2\phi,
\]
which follows from integrating by parts in the action. This trick, inherited from \texttt{LATTICEEASY}, allows reuse of the Laplacian operator used in \cref{eq:fullcont} for evolving the scale factor as well.

To evolve $a(\tau)$ numerically, one of the Friedmann equations must be used. We arbitrarily choose the second equation for time evolution, using the first as an energy conservation check—as done in \texttt{LATTICEEASY}.

	\subsection{Working with a fixed metric}
	Throughout the simulation, the background spacetime is modeled by the FLRW metric $g_{\mu\nu} = a^2(\tau)(-d\tau^2 + d\vec{x}^2)$, without metric perturbations $\delta g_{\mu\nu}$. As far as scalar perturbations are concerned, setting $\delta g_{ij} = 0$ is a gauge choice: we work in a flat spatial gauge. However, the non-dynamical components $\delta g_{0\mu}$ are in principle constrained by the Einstein equations. These are suppressed when $\epsilon = -\dot{H}/H^2 \ll 1$ (with $H$ the Hubble parameter and a dot, e.g. $\dot{H}$, denoting derivative with respect to cosmic time $t$). In this regime, corresponding to the decoupling limit of gravity~\cite{Cheung:2007st, Behbahani:2011it,Creminelli:2024cge}, neglecting $\delta g_{0\mu}$ is a valid approximation.\footnote{Beyond scalar perturbations, a scalar field also sources tensor perturbations at higher orders in perturbation theory (see \cref{sec:GW}). Our simulation neglects these tensor perturbations, assuming they remain decoupled, which is consistent with the decoupling limit of gravity (see~\cite{Caravano:2024moy,Caravano:2025diq} and references therein).}

The validity of this approximation relies on two related conditions: accelerated expansion, quantified by $\epsilon \ll 1$, and a subdominant inhomogeneous contribution to the local energy budget, in particular from spatial gradients of the inflaton.\footnote{When gradients become a sizable fraction of the energy in the system, the notion of a globally defined $\epsilon$ loses meaning, which is why it is important to specify both criteria.} If $\epsilon$ approaches unity, or if gradient energy becomes comparable to the homogeneous background energy, the decoupling picture can fail and local variations of the expansion rate may become relevant. This breakdown is typically associated with the end of inflation: as long as inflation is sustained, the fixed-metric approximation is generally expected to remain valid. This intuition is supported by recent numerical relativity simulations~\cite{Launay:2025kef}, which agree well with lattice results obtained with an FLRW metric~\cite{Caravano:2024tlp}.

Even when $\epsilon \ll 1$, metric perturbations can still induce small but potentially relevant corrections to observables. In such cases, they can be reintroduced perturbatively, as done for example in~\cite{Caravano:2024xsb, Jamieson:2025ngu}. However, \texttt{InflationEasy} does not yet include these corrections.

    \section{Numerical implementation}
    \label{sec:implementation}
    In this section, we provide an overview of the numerical implementation, from the discretization of the equations of motion to the calculation of physical outputs. The content summarized here draws from the developments in~\cite{Caravano:2021pgc,Caravano:2022yyv,Caravano:2024tlp,Caravano:2024moy, Caravano:2025diq, gwpaper, Caravano:2021bfn,Caravano:2022epk,Caravano:2024xsb}, and serves as a general introduction to the tools included in the public release of the code. For additional technical details, we refer the reader to these references.

\subsection{Discretization and lattice equations of motion}
\label{sec:discrete}

To solve \cref{eq:fullcont} numerically, we discretize space on a cubic lattice. The lattice consists of $N_{\rm pts}^3$ points, separated by a comoving lattice spacing $\Delta x = L / N_{\rm pts}$, where $L$ is the comoving physical length of the simulation box. A continuous field $f(\vec{x})$ is mapped to a discrete array $f(\vec{n})$:
\begin{equation}
\label{eq:discretization}
f(\vec{x}), \quad \vec{x} \in \mathbb{R}^3 \quad\longrightarrow\quad f(\vec{n}), \quad \vec{n} \in \mathbb{N}^3,\quad n_i \in \{1, \dots, N_{\rm pts}\}.
\end{equation}
We impose periodic boundary conditions so that, for example, $f(N_{\rm pts}, n_2, n_3) = f(1, n_2, n_3)$. Under this discretization, the partial differential equation \cref{eq:fullcont} becomes a system of $N_{\rm pts}^3$ coupled ordinary differential equations:
\begin{equation}
\label{eq:disc}
\partial_\tau^2 \phi(\vec{n}) + 2\mathcal{H}\partial_\tau \phi(\vec{n}) - [\nabla^2 \phi](\vec{n}) + a^2 \frac{\partial V}{\partial \phi}(\vec{n}) = 0.
\end{equation}
The coupling arises through the discrete Laplacian $[\nabla^2 \phi](\vec{n})$, which we define using a second-order central finite-difference stencil~\cite{press1986numerical}:
\begin{equation}
\label{eq:discretelaplacian}
[\nabla^2 f](\vec{n}) = \frac{1}{(\Delta x)^2} \sum_{\alpha = \pm 1} \left(
f(\vec{n} + \alpha \vec{e}_1) + f(\vec{n} + \alpha \vec{e}_2) + f(\vec{n} + \alpha \vec{e}_3) - 3f(\vec{n})
\right),
\end{equation}
where $\vec{e}_1 = (1, 0, 0)$, $\vec{e}_2 = (0, 1, 0)$, and $\vec{e}_3 = (0, 0, 1)$. This approximation converges to the continuum Laplacian as $\Delta x \rightarrow 0$, with a truncation error of order $O(\Delta x^2)$. While higher-order stencils can be implemented for greater accuracy (as discussed in Appendix C of~\cite{Caravano:2021pgc}), the present public release includes only the second-order scheme.

\subsubsection{Modified dispersion relation}
\label{sec:modified}

Discretizing space alters the dynamics of waves on the lattice compared to continuous space. To see this, consider the wave equation for a free scalar field:
\begin{equation}
\partial_\tau^2 \psi(\vec{x}, \tau) = \nabla^2 \psi(\vec{x}, \tau) \quad \xrightarrow{\text{FT}} \quad \partial_\tau^2 \psi(\vec{k}, \tau) = -k^2 \psi(\vec{k}, \tau),
\end{equation}
where FT denotes a Fourier transform. In discrete space, the finite-difference Laplacian yields a different evolution in Fourier space:
\begin{equation}
\partial_\tau^2 \psi(\vec{n}, \tau) = [\nabla^2 \psi](\vec{n}, \tau) \quad \xrightarrow{\text{DFT}} \quad \partial_\tau^2 \psi(\vec{\kappa}, \tau) = -k_{\rm eff}^2(\vec{\kappa}) \psi(\vec{\kappa}, \tau),
\end{equation}
where $\vec{\kappa}$ is the discrete lattice momentum,
\begin{equation}
\vec{\kappa} = \frac{2\pi}{L} \vec{m}, \quad \vec{m} \in \mathbb{N}^3, \quad m_i \in \{1, \dots, N_{\rm pts}\},
\end{equation}
and the effective wavenumber $k_{\rm eff}$ is given by
\begin{equation}
\label{eq:eff}
\vec{k}_{\rm eff}(\vec{\kappa}) = \frac{2N_{\rm pts}}{L} \sin\left( \frac{\pi \vec{m}}{N_{\rm pts}} \right).
\end{equation}
This defines a modified dispersion relation that converges to the continuum form $k^2$ only in the infrared: $k^2_{\rm eff}(\vec{\kappa}) \to {\kappa}^2$ when $m_i \ll N_{\rm pts}$. Thus, wave propagation on the lattice differs from that in continuous space due to the discreteness of the grid.

This issue can be resolved by identifying $\vec{k}_{\rm eff}(\vec{\kappa})$ with the physical momentum $\vec{k}$ of the corresponding continuum mode, instead of $\vec\kappa$:
\begin{equation}
\vec{k}_{\rm eff}(\vec{\kappa}) \longleftrightarrow \vec{k}.
\end{equation}
This prescription, introduced in~\cite{Caravano:2021pgc}, makes the lattice evolution equivalent to continuum evolution. While we illustrated this equivalence for a free wave equation, it has also been shown to hold for inflationary dynamics in single-field models~\cite{Caravano:2021pgc,Caravano:2022yyv}. This is not true in general—for instance, in axion-gauge models such as axion-$U(1)$ inflation, this prescription is insufficient~\cite{Caravano:2021bfn,Caravano:2022epk}, and one must resort to higher-order schemes or Fourier-based differentiation.

In \texttt{InflationEasy}, this effective wavenumber prescription is used to correct for discretization both in setting the initial conditions and in computing output observables such as the inflaton power spectrum. As a result, outputs such as power spectra become insensitive to the choice of lattice spacing, making this one of the key improvements over other lattice codes.

\subsubsection{Equations in code units}

The code works in rescaled spatial and temporal coordinates $(\tilde{x}, \tilde{\tau})$, defined via
\begin{equation}
\label{eq:rescaling}
d\tau \rightarrow d\tilde{\tau} = B a^s\, d\tau, \quad x \rightarrow \tilde{x} = Bx,
\end{equation}
where $B$ is a constant with dimensions of inverse length and $s$ is a dimensionless scaling parameter. This rescaling is inspired by the conventions of \texttt{LATTICEEASY}, and enhances the numerical stability of the equations.

In these rescaled variables, the equations of motion take the form:
\begin{align}
\label{eq:eoms}
\begin{split}
&\partial_{\tilde{\tau}}^2 \phi + (2 + s)\frac{\partial_{\tilde{\tau}} a}{a} \, \partial_{\tilde{\tau}} \phi - a^{-2s} [\tilde{\nabla}^2 \phi] + a^{2 - 2s} \frac{\partial \tilde{V}}{\partial \phi} = 0,\\
&\partial_{\tilde{\tau}}^2 a = -s \frac{(\partial_{\tilde{\tau}} a)^2}{a} + \frac{1}{3} a^{-2s + 3} \left( \langle \tilde{\rho} \rangle - 3 \langle \tilde{p} \rangle \right),
\end{split}
\end{align}
where all quantities with tildes have been normalized by appropriate powers of $B$ (e.g., $\tilde{V} = V / B^2$). The spatial derivatives $[\tilde{\nabla}^2 \phi]$ are computed using the same finite-difference stencil described earlier, but now expressed in terms of the rescaled lattice spacing $\tilde{\Delta x} = B\Delta x$.

Internally, all evolution is carried out in these code units. However, to improve interpretability, all output quantities of the code (e.g., power spectra, background observables) are automatically converted to reduced Planck units.

\subsection{Initial Conditions}
\label{sec:ic}

We now describe the generation of initial conditions on the lattice. While the procedure follows the standard approach used in most lattice simulations, \texttt{InflationEasy} includes one significant improvement: the initial fluctuations are generated in a way that accounts for the modified dispersion relation induced by the lattice, as discussed in \cref{sec:modified}. This feature, first introduced in~\cite{Caravano:2021pgc}, ensures consistency between the initialization and the discrete dynamics.

Although the evolution of the field is fully non-perturbative—with no background/perturbation split introduced—the initial conditions do rely on such a separation. Specifically, the field is initialized as a classical background plus quantum fluctuations. We now describe the initialization of these two components.

\subsubsection{Background values}

The initial background values of the inflaton, $\bar{\phi}(\tilde\tau_{\rm in})$, and its velocity, $\partial_{\tilde\tau} \bar{\phi}(\tilde\tau_{\rm in})$, are set by the user in the run-time parameter file \texttt{params.txt} (or, if absent, by defaults compiled in \texttt{runtime\_parameters.cpp}). A typical choice, adopted in the default example (see \cref{sec:default}), is to set them according to the homogeneous inflationary trajectory—i.e., the solution to \eqref{eq:fullcont} with the gradient term omitted. This simple ordinary differential equation can be solved starting a few $e$-folds before the simulation start time, in order to determine the precise value of $\partial_{\tilde\tau} \bar{\phi}(\tilde\tau_{\rm in})$ corresponding to the chosen $\bar{\phi}(\tilde\tau_{\rm in})$.

The scale factor is initialized as $a = 1$, and its time derivative in program units is determined from the Friedmann equation using the background energy density:
\begin{equation}
\label{eq:initial_H}
3\,\big[\partial_{\tilde\tau} a(\tilde\tau_{\rm in})\big]^2 = \bar\rho = \frac{\big(\partial_{\tilde\tau} \bar\phi(\tilde\tau_{\rm in})\big)^2}{2} + V(\bar\phi).
\end{equation}
After generating the fluctuations (as described below), the initial value of $\partial_{\tilde\tau} a$ is updated to include the contribution from the gradient energy, computed as a lattice average of $(\vec{\tilde\nabla} \phi)^2 / 2$. While quantum vacuum fluctuations should not source the background equations, their inclusion here reflects the semi-classical approximation used in the lattice approach. Importantly, omitting this gradient term would introduce a residual curvature term into the second Friedmann equation, compromising the consistency of the evolution. Including it also enables accurate monitoring of energy conservation during the simulation, as discussed in \cite{latticeeasy,Caravano:2022yyv}.

\subsubsection{Fluctuations}

We now describe the initialization of field fluctuations. A detailed derivation can be found in~\cite{Caravano:2022yyv}, but we summarize the main steps here.

We begin by defining the discrete version of the Bunch-Davies vacuum\footnote{Strictly speaking, the Bunch–Davies vacuum is defined in exact de Sitter spacetime, whereas in quasi-de Sitter backgrounds, the vacuum is typically referred to as ``adiabatic vacuum''.}:
\begin{equation}
\label{eq:discBD}
u(\vec\kappa) = \frac{L^{3/2}}{a \sqrt{2 \omega_{\vec\kappa}}} e^{-i \omega_{\vec\kappa} \tau}, \quad\quad \omega_{\vec\kappa}^2 = k_{\rm eff}^2(\vec\kappa) + m^2 \simeq k_{\rm eff}^2(\vec\kappa).
\end{equation}
In this expression, we have neglected the effective mass-squared of the inflaton fluctuations, $m^2 = \frac{\partial^2 V}{\partial\phi^2}$. This is typically a safe assumption, as $m^2$ is negligible compared to $k_{\text{eff}}^2$ sufficiently inside the horizon. The effective mass of the inflaton can easily be included by modifying this expression in the file \texttt{initialize.cpp}. The normalization factor $L^{3/2}$ accounts for the finite volume of the lattice box (see~\cite{Caravano:2022yyv} for details). Crucially, this mode function incorporates the lattice-corrected dispersion relation via $k_{\rm eff}$, as defined in \cref{eq:eff}, in contrast to standard approaches which use the lattice momenta $\vec{\kappa} = 2\pi \vec{m} / L$.

To generate classical field fluctuations, we assign each Fourier mode a random complex amplitude drawn from a Rayleigh distribution with variance $|u(\vec\kappa)|^2$:
\begin{equation}
\label{eq:randommodes}
\phi(\vec\kappa) = e^{i 2\pi Y_{\vec{\kappa}}} \sqrt{-\ln(X_{\vec{\kappa}})} \, u(\vec\kappa),
\end{equation}
where $X_{\vec{\kappa}}$ and $Y_{\vec{\kappa}}$ are independent random variables uniformly distributed in the interval $[0, 1]$. 

An inverse discrete Fourier transform is then applied to obtain the real-space field configuration. The fluctuations are added to the background value $\bar{\phi}_{\rm in}$ to form the initial condition for the inflaton field. The initial values of the field’s time derivative are generated similarly, using the time derivative of the mode functions $\partial_{\tilde\tau} u(\vec\kappa_{\vec{m}})$ and the same random realizations. Note that we are not using the same procedure of \texttt{LATTICEEASY}, which is known to suffer from a subtle inconsistency in the treatment of the field and its conjugate momentum~\cite{Frolov_2008}.

\subsection{Time integration and \texorpdfstring{$\delta N$}{deltaN} calculation}

Once the initial conditions are specified, the system evolves under the equations of motion \cref{eq:eoms}. The default option is the same as in \texttt{LATTICEEASY}: a second-order symplectic leapfrog scheme in which the field $\phi(\vec{n})$ and its velocity $\partial_{\tilde\tau} \phi(\vec{n})$ are stored at half-step offsets in time. Because leapfrog is standard, we refer the reader to the documentation of \texttt{LATTICEEASY}~\cite{latticeeasy} for details. The code also supports RK4 and adaptive RK45 integrators. The integrator can be selected in \texttt{params.txt}.

In leapfrog and RK4 modes, the baseline step is rescaled dynamically as
\begin{equation}
\Delta \tilde{\tau} = \Delta \tilde{\tau}_{\rm in} \, a^{s-1},
\end{equation}
which corresponds to a constant timestep in cosmic time $t$. In RK45 mode, this quantity is used as a reference scale for the adaptive controller, while the accepted step is adjusted by the local error estimate and constrained by the RK45 run-time bounds.

The simulation stops when the scale factor reaches a final value set by the user in the configuration file (see \cref{sec:configuring}). At this point, \texttt{InflationEasy} outputs statistical summaries of the simulation. After the main evolution, the code optionally performs a $\delta N$ calculation to extract the comoving curvature perturbation $\zeta$. Formally, $\delta N$ is defined between an initial flat slice and a final uniform-total-energy slice. In the implementation, this final slice is approximated by the local time delay in reaching a reference field value $\phi_{\rm ref}$:
\begin{equation}
\zeta(\vec{x}) = N(\vec{x}) - \langle N(\vec{x}) \rangle,
\end{equation}
where $N(\vec{x})$ is the number of e-folds experienced at each point until the field reaches $\phi_{\rm ref}$. To compute $N(\vec{x})$, the code switches to a ``separate universe'' evolution, where each lattice point evolves independently with gradient terms neglected. This is a valid approximation when the lattice grid spacing $\Delta x$ has become larger than the horizon.

To perform this calculation, the time variable is converted to e-folds time, $dN = H dt$, and the evolution follows \cite{Caravano:2024moy,Caravano:2025diq}:
\begin{align}
\label{eq:eom_deltaN}
\frac{d^2\phi}{dN^2}(\vec{n}) + \left[3 - \frac{1}{2}\left(\frac{d\phi}{dN}(\vec{n})\right)^2 \right]
\left[ \frac{d\phi}{dN}(\vec{n}) + \frac{V_{,\phi}(\phi)}{V(\phi)}(\vec{n}) \right] = 0.
\end{align}
This equation is integrated until each point reaches the reference value $\phi_{\rm ref}$ according to the stopping criterion used by the code. In practice, the code needs a simple indication of how the potential behaves near the end of the run: monotonic, anti-monotonic, or neither. If this behavior is known, the stopping test can be done with a faster field-based criterion; if not, the code can use a fully generic potential-based criterion. Concretely, the implemented conditions are monotonic potential ($|\phi|>|\phi_{\rm ref}|$ during evolution), anti-monotonic potential ($|\phi|<|\phi_{\rm ref}|$ during evolution), and generic fallback ($V(\phi)>V(\phi_{\rm ref})$ during evolution). Accordingly, the automatic choice of $\phi_{\rm ref}$ corresponds to the final-box point minimizing the same criterion (for the generic potential criterion: minimum $V$), rather than explicitly minimizing the full local energy density $K+G+V$. This is expected to be a good approximation when the run ends in a separate-universe, near-attractor regime in which gradients are small and $\dot\phi$ is effectively determined by $\phi$, so constant-$\phi$ slices approximately track constant-energy slices. In non-attractor regimes, this equivalence can break down, and one should test the stability of $\zeta$ against the choice of $\phi_{\rm ref}$ and \texttt{Nend} (or set \texttt{phiref\_manual} explicitly)~\cite{Caravano:2025diq}.

For further details on the $\delta N$ approach, including the distinction between constant field vs. constant energy hypersurfaces, see \cite{Caravano:2025diq}.
\subsection{Outputs}
\label{sec:output}

 \begin{figure}[htbp]
  \centering
  \scalebox{0.92}{
  \begin{tikzpicture}
    \def\xA{0}
    \def\xB{7.3}  

    \def\yA{0}
    \def\yB{-6.5}
    \def\yC{-13}

    \node at (\xA,\yA) {\includegraphics[width=7.5cm]{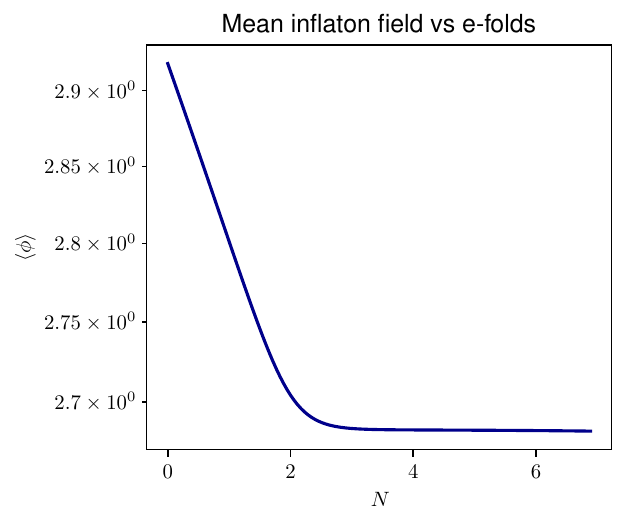}};
    \node at (\xB,\yA) {\includegraphics[width=7cm]{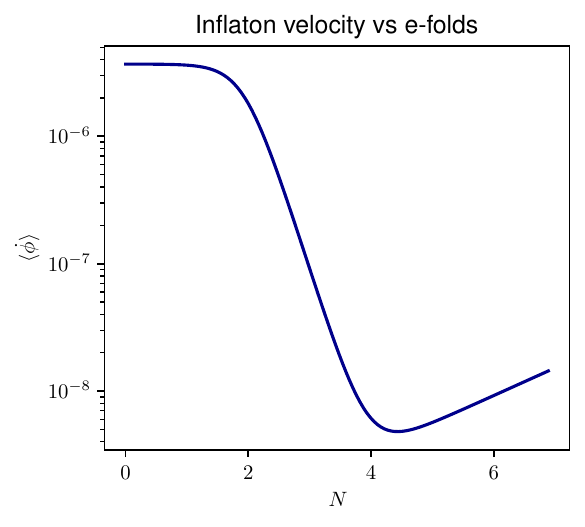}};

    \node at (\xA+0.2,\yB) {\includegraphics[width=7cm]{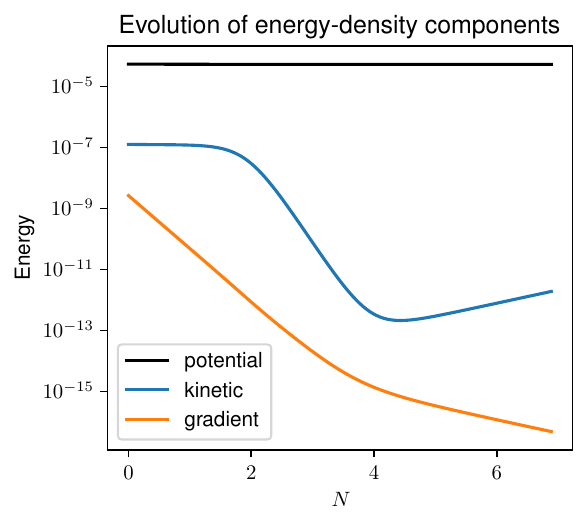}};
    \node at (\xB,\yB) {\includegraphics[width=7cm]{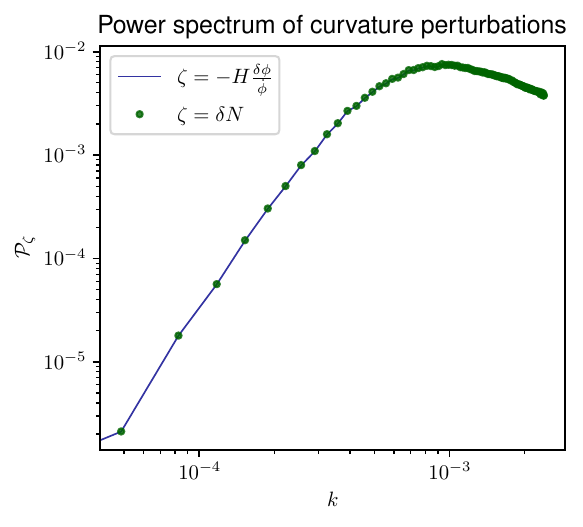}};

    \node at (\xA+0.2,\yC) {\includegraphics[width=6.9cm]{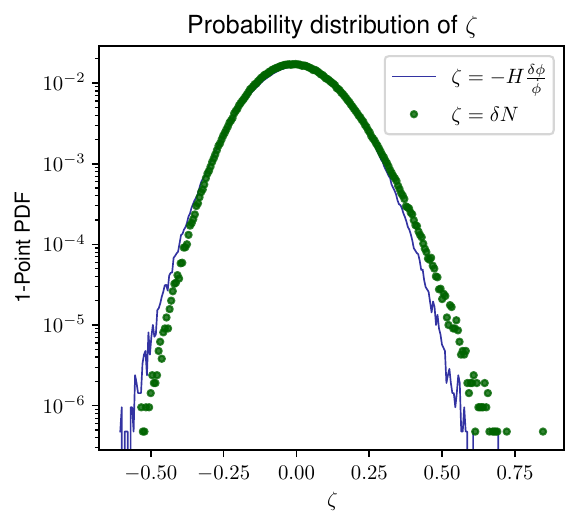}};
    \node at (\xB+0.4,\yC) {\includegraphics[width=5cm]{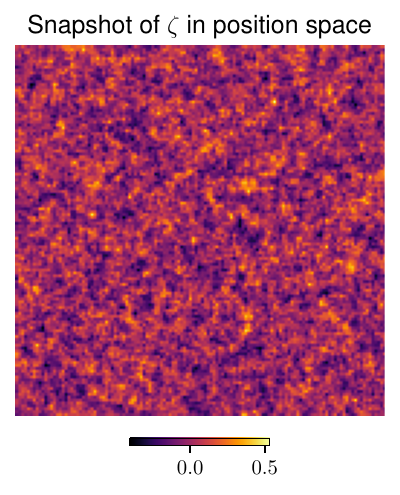}};
  \end{tikzpicture}
  }
  \caption{Example of typical outputs from the lattice simulation. \textbf{Top:} Evolution of the spatial average of the inflaton field and its velocity as a function of the number of $e$-folds $N$. \textbf{Middle:} Left panel shows the evolution of the different energy components in the lattice; right panel displays the power spectrum of the curvature perturbation $\zeta$, computed both via the linear relation and the $\delta N$ method. \textbf{Bottom:} Left panel shows the probability distribution function (PDF) of $\zeta$; right panel shows a 2D slice of the final $\zeta$ field in position space. These outputs correspond to the default example in the code—namely, the ultra-slow-roll (USR) potential producing a peak in the power spectrum—run on a lattice with $N^3_{\rm pts} = 128^3$ points. These plots are generated using the example notebook \texttt{{plot.ipynb}}, included in the public release under the \texttt{notebooks/} directory.}
  \label{fig}
\end{figure}

We now describe the various outputs produced by the code, providing a brief overview of the physical quantities computed during the simulation. A more detailed discussion can be found in~\cite{Caravano:2022yyv}. The key distinction between \texttt{InflationEasy} and other lattice codes is that Fourier-based outputs—such as the inflaton and curvature power spectra—account for the modified dispersion relation. This ensures that the results are independent of the lattice implementation and can be directly compared to analytical predictions. 

All output files are saved in the \texttt{{results/}} directory in \texttt{.dat} format. The public release of the code includes a Python notebook, \texttt{{plot.ipynb}}, which provides examples of how to load and visualize these outputs. Note that not all quantities are computed by default; the generated outputs are controlled primarily via \texttt{params.txt}. In \cref{fig}, we show examples of outputs produced by the lattice code in the default setup, which will be discussed in detail below and further in \cref{sec:instructions}. In this section we focus on scalar-sector outputs. Gravitational-wave outputs are documented separately in \cref{sec:GW}.

The following files may be generated:

\begin{itemize} 
    \item \textbf{Background files:}\\ \texttt{means.dat}, \texttt{velocity.dat}, \texttt{variance.dat} and \texttt{sf.dat}. The first three contain, respectively, the mean inflaton field $\langle \phi \rangle$, its velocity $\langle \dot\phi \rangle$, and its variance $\langle \phi^2 \rangle - \langle \phi \rangle^2$, computed as averages over the $N^3_{\rm pts}$ lattice points. Files \texttt{means.dat}, \texttt{velocity.dat}, and \texttt{variance.dat} each have three columns: simulation time $\tilde{\tau}$, scale factor $a$, and the corresponding quantity. The file \texttt{sf.dat} has four columns: simulation time $\tilde{\tau}$, scale factor $a$, Hubble rate $H=\dot{a}/a$, and $\ddot a$.

    \item \textbf{Energy files:} \\ \texttt{energy.dat} and \texttt{conservation.dat}. The file \texttt{energy.dat} contains the energy budget with the following columns: simulation time $\tilde{\tau}$, scale factor $a$, kinetic energy, gradient energy, and potential energy. The file \texttt{conservation.dat} provides a check of energy conservation, using the Friedmann equation. Its columns are: simulation time $\tilde{\tau}$, scale factor $a$, and the quantity $E = \frac{3\mathcal{H}^2}{\rho a^2}$.

    \item \textbf{Inflaton spectra files:} \\ \texttt{spectra.dat} and \texttt{spectratimes.dat}. The former contains the inflaton power spectrum stored as a concatenated list of spectra at different times, separated by empty lines. The latter lists the times at which the spectra are computed. Additionally, the file \texttt{modes.dat} contains the comoving effective momenta at which these spectra are calculated, calculated as binned mean of the expression \cref{eq:eff} on the same bins used for the spectra. These comoving modes are given in Planck units (see the discussion after \cref{eq:modes} for how to convert them to physical units).

    \item \textbf{Curvature perturbation spectra:} \\ \texttt{spectraN.dat} and \texttt{spectraLOG.dat}. These contain the power spectrum of $\zeta$, calculated using the non-perturbative $\delta N$ method, and a log-based approximation:
    \begin{equation}
        \zeta_{\rm log} = \frac{1}{\eta} \log\left(1 - \eta\, \delta\phi\, \frac{H}{\langle \dot\phi \rangle} \right),
    \end{equation}
    which assumes a constant second slow-roll parameter $\eta$ (see \cite{Pi:2022ysn,Caravano:2025diq}). The value of $\eta$ must be set in \texttt{params.txt}. These spectra are evaluated only at the final simulation time, in contrast to the inflaton spectra.

    \item \textbf{Histogram files:} \\ \texttt{histogram.dat}, \texttt{histogramN.dat}, \texttt{histogramLOG.dat}, along with \texttt{histogramtimes.dat},\\\texttt{histogramtimesN.dat}, and \texttt{histogramtimesLOG.dat}. The first three files contain binned histograms of the inflaton field and the curvature perturbation $\zeta$, using either the full $\delta N$ result or the log-based approximation. The corresponding \texttt{histogramtimes} files include the simulation times at which histograms were computed, along with binning data required to reconstruct the $x$-axis of the histograms.

    \item \textbf{Snapshot files:} \\ \texttt{snapshots\_2d\_phi.dat}, \texttt{snapshots\_2d\_phidot.dat}, and \texttt{snapshots\_2d\_deltaN.dat}. These files contain 2D slices of the inflaton field, its velocity, and $\zeta=\delta N$. The first two are recorded at multiple times during the simulation, while the $\delta N$ snapshot is saved only at the final time. Optionally, the code can also produce 3D snapshots of $\phi$, although this is memory intensive.
\end{itemize}

    \subsection{Code structure}
    \label{sec:structure}
   The source code is contained in the directory \texttt{src/}, which includes the following files:
\begin{itemize}
\item \texttt{parameters.h} – Compile-time configuration header (feature toggles, lattice size, and preprocessor branches). Changes here require recompilation.
\item \texttt{runtime\_parameters.cpp} – Defines default run-time values and the parser for \texttt{params.txt}.
\item \texttt{potential.cpp} – Contains all functions related to the potential. To change the analytical form of the potential, users should modify the first two functions in this file.
\item \texttt{main.cpp} – Contains the entry point of the program and coordinates the main evolution steps.
\item \texttt{main.h} – Header file associated with \texttt{main.cpp}, declaring key functions and including necessary headers.
\item \texttt{initialize.cpp} – Responsible for initializing the field configurations, including setting initial conditions and allocating memory.
\item \texttt{evolution.cpp} – Implements the main evolution loop, computing the time evolution of fields and their derivatives using finite-difference methods.
\item \texttt{output.cpp} – Handles all output-related tasks, including writing data to disk and computing summary statistics at given time steps.
\end{itemize}

\section{Usage Instructions}
\label{sec:instructions}
\subsection{Downloading the Code}
The code is available on GitHub at \href{https://github.com/caravangelo/inflation-easy.git}{https://github.com/caravangelo/inflation-easy.git}.  
It can be downloaded in different ways. For example, users can download a ZIP archive directly from the GitHub page, or clone the repository with \texttt{git}.  
If using the terminal, navigate to the directory where you want to place the code and run:
\begin{lstlisting}[language=bash,numbers=none]
git clone https://github.com/caravangelo/inflation-easy.git
\end{lstlisting}

\subsection{Quickstart Notebook}

For first-time users, especially non-coders, we recommend starting from the Jupyter quickstart notebook \texttt{notebooks/quickstart.ipynb}. It provides a guided end-to-end workflow to compile and run the code with a lightweight default setup (analytical potential, small $64^3$ lattice). The notebook stores outputs in a dedicated folder, produces summary plots automatically, and clearly separates compile-time and run-time settings.

\subsection{Compiling and Running the Code}

To run the code, you first need to compile it. Open a terminal, navigate to the code directory, and run:
\begin{lstlisting}[language=bash,numbers=none]
make
\end{lstlisting}
This command builds the executable using the compiler selected by the \texttt{Makefile} and its build flags. Once compiled, run the simulation using:
\begin{lstlisting}[language=bash,numbers=none]
./inflation_easy
\end{lstlisting}
This runs the simulation with the default configuration. A ready-to-use \texttt{params.txt} is included in the repository root; users can edit its key-value entries to override run-time defaults. Output files (described in \cref{sec:output}) will be saved in the \texttt{results/} directory. During execution, the code prints the current scale factor and number of time steps to the terminal; the print cadence is controlled by \texttt{output\_freq} in \texttt{params.txt}. The parameter \texttt{output\_infrequent\_freq} controls the cadence of the heavier diagnostic outputs.

A Python notebook \texttt{plot.ipynb} is included in the \texttt{notebooks/} directory. It provides instructions for loading and plotting quantities such as the power spectrum and field histograms (including the ones of $\zeta$) at the end of the simulation. In \cref{fig}, we show example outputs from the notebook corresponding to the default code example, which will be introduced in the next section.

\subsection{Default Options}
\label{sec:default}

The code includes two default potential options:

\begin{itemize}
    \item \textbf{Ultra-slow-roll potential}: A numerical potential featuring an inflection point that generates a pronounced peak in the primordial power spectrum. This corresponds to one of the Case I potentials in \cite{Caravano:2024moy, Caravano:2025diq}, specifically the one with a tree-level peak amplitude of the power spectrum\footnote{In \cite{Caravano:2024moy,Caravano:2025diq}, we classify the potentials using the tree-level power spectrum, i.e., the one predicted by linear perturbation theory.} of $\mathcal{P}_{\zeta,\rm tree}^{\rm peak} \sim 10^{-2}$. Concretely, this potential is obtained by prescribing a phenomenological SR--USR--SR profile for $\eta(N)$ and numerically reconstructing $V(\phi)$ \cite{Franciolini:2022pav}. Therefore the default USR potential is provided as tabulated files in \texttt{inputs/}, rather than as a closed analytic expression.

    \item \textbf{Hilltop potential}: An analytic hilltop potential \cite{Boubekeur:2005zm} given by \begin{equation}
    V(\phi) = V_0 \left(1 - \frac{1-n_s}{2}\frac{\phi^2}{2 M_{\rm Pl}^2}\right).
    \end{equation}
\end{itemize}
By default, the code runs with the numerical USR potential on a lattice of size $N_{\rm pts}^3 = 128^3$ and runs until the scale factor reaches $a=2N_{\rm pts}=256$ (where $a=1$ at the beginning of the simulation), corresponding to $N = \log(256) \simeq 5.5$ e-folds of evolution. The total number of $e$-folds must be chosen based on the spatial resolution, which ultimately dictates the maximum running time of the simulation. See the next section for how to set the total number of e-folds for a given lattice size.

In \cref{fig}, we show the outputs of the code for the default example with a USR potential. These plots are generated using the example Python notebook \texttt{plot.ipynb} located in the \texttt{notebooks/} directory. They display lattice outputs such as background quantities $\langle \phi \rangle$ and $\langle \dot{\phi} \rangle$, energy components, the power spectrum of $\zeta$, its 1-point probability distribution function (PDF), and a 2D snapshot of $\zeta$ in real space.

\subsubsection*{Running time}

For the numerical USR default example distributed with this release, the following wall-clock times provide a reference benchmark for a $128^3$ lattice. Measurements were obtained on an Apple M3 Max (14 CPU cores, 36\,GB RAM), macOS 26.3, compiled with \texttt{Homebrew clang 21.1.2}, with OpenMP enabled and \texttt{OMP\_NUM\_THREADS=14}:
\begin{itemize}
    \item inflation only simulation, no $\delta N$ calculation, no GWs: $183.49\,\mathrm{s}$ ($3$ min $3$ s);
    \item inflation + $\delta N$, no GWs: $248.52\,\mathrm{s}$ ($4$ min $9$ s);
    \item inflation only, no $\delta N$, with inflationary GWs: $470.45\,\mathrm{s}$ ($7$ min $50$ s);
    \item full run (inflation + $\delta N$ + post-inflation + inflationary and post-inflationary GWs): $577.65\,\mathrm{s}$ ($9$ min $38$ s).
\end{itemize}

\subsubsection*{Switching}
Switching between the two defaults requires two explicit steps.
This split is intentional: the two cases correspond to different physics inputs (tabulated numerical potential versus analytic potential), so they use different code paths and run-time parameter presets. \vspace{0.5mm} \\ 
\textbf{Step 1 (compile-time choice).} In \texttt{parameters.h}, choose which potential type to compile:

\begin{lstlisting}[language=C++, firstnumber=3,caption={},numbers=none]
// Set to 1 to enable a numerical potential loaded from file, 0 for analytic expression
#define numerical_potential 1
\end{lstlisting}
Because this is a compile-time flag, you must recompile after changing it. \vspace{0.5mm} \\ 
\textbf{Step 2 (run-time parameter set).} Match \texttt{params.txt} to the selected potential type. The repository provides two presets: \texttt{params.numerical.txt} and \texttt{params.analytic.txt}. Copy the appropriate one to \texttt{params.txt}:
\begin{itemize}
    \item \textbf{Analytic default}: set \texttt{numerical\_potential} to \texttt{0}. Copy \path{params.analytic.txt} to \path{params.txt}, then rebuild.
    \item \textbf{Numerical default}: set \texttt{numerical\_potential} to \texttt{1}. Copy \path{params.numerical.txt} to \path{params.txt}, then rebuild.
\end{itemize}

\subsection{Configuring a New Potential}
\label{sec:configuring}

To use a different potential, follow these two steps:

\subsubsection*{Step 1: Define the Potential}

For an analytic potential, edit the following functions in \texttt{potential.cpp}:

\begin{lstlisting}[language=C++, firstnumber=14,caption={}]
double analytic_potential(double field_value) {
    return V0 * (1. - (1. - ns) / 4. * pw2(field_value)) / pw2(rescale_B);
}

double analytic_potential_derivative(double field_value) {
    return -V0 * (1. - ns) / 2. * field_value / pw2(rescale_B);
}
#endif
\end{lstlisting}
These return the potential and its derivative. Modify them as needed for your model.

For a numerical potential, update the files in \texttt{inputs/}: 
\begin{itemize}
    \item \texttt{field\_values.dat}: field points $\phi$ at which $V(\phi)$ is evaluated
    \item \texttt{potential.dat}: values of $V(\phi)$
    \item \texttt{potential\_derivative.dat}: values of $V'(\phi)$
\end{itemize}
All files must be single-column. Note that running simulations with an analytic potential is much faster than interpolating a numerical potential.

\subsubsection*{Step 2: Set Run-time Parameters in \texttt{params.txt}}

Once the potential is defined, the next step is to configure the relevant run-time parameters in \texttt{params.txt}. The compile-time header \texttt{parameters.h} should be used only for options that affect compiled code paths (feature toggles and lattice size).

The first parameters to set are the physical parameters of the potential. For example, for the hilltop potential, one of the default models, this reads:
\begin{lstlisting}[language=bash,numbers=none]
V0 = 3.338e-13
ns = 0.97
\end{lstlisting}
This defines $V_0$, the overall amplitude of the potential, and the scalar spectral index $n_s$. If your model includes more than one parameter (e.g. a coupling or a second mass scale), you should declare each of them in this section.

Next, set the field rescaling factor, referred to as $B$ in \cref{eq:rescaling}:
\begin{lstlisting}[language=bash,numbers=none]
# Optional: if omitted, the code sets rescale_B = sqrt(V0)
rescale_B = 5.778e-7
\end{lstlisting}
This parameter sets the code units for both the time derivatives and the lattice spacing. A typical and convenient choice is to use either the value of the Hubble rate or the inflaton mass at the beginning of the simulation. All dimensionful quantities in the simulation are rescaled by this factor, so this choice determines the unit system in which the simulation runs.

After the physical parameters and code units are defined, you must set the initial value of the homogeneous part of the inflaton field and its time derivative in code units:
\begin{lstlisting}[language=bash,numbers=none]
initial_field = 0.0935
initial_derivative = 0.000796
\end{lstlisting}
These values determine the starting point of the background evolution and should be chosen based on where you want the simulation to begin along the inflationary trajectory. 

You must also choose the comoving box size of the lattice:
\begin{lstlisting}[language=bash,numbers=none]
L = 10.0
\end{lstlisting}
The comoving size $L$, together with the number of lattice points $N_{\rm pts}$ (set at compile time in \texttt{parameters.h}), determines the infrared and ultraviolet cutoffs of the simulation \cite{Caravano:2021pgc}:\footnote{
To express these comoving momenta in physical units (e.g.\ ${\rm Mpc}^{-1}$), two steps are required. First, one converts from reduced Planck units to the desired physical unit system using standard unit conversions. Second, the comoving momenta are defined with respect to the scale factor normalization used in the simulation; converting them to comoving momenta defined at a later reference time (such as today) requires specifying the post-inflationary expansion history, including assumptions about reheating and the subsequent radiation- and matter-dominated eras.
}

\begin{equation}
\label{eq:modes}
k_{\rm IR} = \frac{2\pi}{L},\quad\quad k_{\rm UV} = \frac{2\sqrt{3}}{L}N_{\rm pts}.
\end{equation}

To ensure accurate initial conditions, the comoving box size should be smaller than the Hubble radius at the start of the simulation, i.e., $L \lesssim 1/(aH)$. A common choice is $L \sim 1/(aH)$, which guarantees that most modes begin well inside the horizon, avoiding unnecessarily long simulations. Note that in this case, the lowest-$k$ mode ($k_{\rm IR}$) may not be accurately initialized. However, the rest of the spectrum is well resolved, and this typically has negligible impact on the overall evolution.

The value of $N_{\rm pts}$, which sets the lattice resolution and UV cutoff, must be balanced against computational cost. Larger values of $N_{\rm pts}$ yield higher resolution, but increase the memory and runtime requirements. For example, simulations larger than $N^3_{\rm pts} = 128^3$ may be difficult to run on a laptop.

The final simulation time can be set by specifying the final value of the scale factor:
\begin{lstlisting}[language=bash,numbers=none]
af = 256
\end{lstlisting}
This entry is optional. If \texttt{af} is not set in \texttt{params.txt}, the code uses the default
\[
a_f = 2N_{\rm pts},
\]
where $N_{\rm pts}$ is the compile-time lattice size set in \texttt{parameters.h}. The chosen value must be large enough to ensure the comoving box is significantly larger than the horizon by the end of the simulation. A good rule of thumb is to stop when the comoving lattice spacing satisfies $\Delta x \sim 1/(aH)$, so that each lattice point evolves as an effectively separate universe. Extending the simulation significantly beyond this time is typically unnecessary and does not affect the dynamics: at that point, no new modes enter the horizon due to the finite UV resolution of the lattice. This remains true unless the chosen potential exhibits interesting super-horizon dynamics, in which case it may be worth running the simulation for a longer time. In practice, one can experiment to find a suitable \texttt{af} value by ensuring that the fluctuation dynamics have frozen by the end of the run (for example, by checking that the power spectrum of the linear curvature perturbation $\zeta_{\rm lin} = -H \delta\phi / \dot\phi$ has stabilized).

As part of a convergence test, it is recommended to vary the values of $L$ and $N_{\rm pts}$ to verify that the physical results do not depend on the lattice cutoffs. This is a standard consistency check in lattice simulations. Note, however, that in some models (particularly those with scale dependence), physical quantities may naturally depend on the cutoffs. In those cases, the goal is not to eliminate cutoff dependence but to determine whether it is a physical feature or a numerical artifact.

Another crucial parameter is the time step for the numerical integration:

\begin{lstlisting}[language=bash,numbers=none]
dt = 0.001
\end{lstlisting}

The time step must be small enough to ensure stability and accuracy. A good practice is to rerun the simulation with different values of \texttt{dt} and check that the results remain unchanged. These convergence tests can be performed, at first, on small grids (e.g., $64^3$) to save time in determining a reasonable range of $dt$ values. If the adaptive RK45 integrator is selected, this fixed \texttt{dt} choice is effectively overridden by the RK45 step-control parameters.

Integrator choices and RK45 controls are specified at run time as
\begin{lstlisting}[language=bash,numbers=none]
inflation_integrator = leapfrog
deltaN_integrator = leapfrog
post_inflation_integrator = leapfrog

rk45_abs_tol = 1e-8
rk45_rel_tol = 1e-6
rk45_min_dt = ...
rk45_max_dt = ...
rk45_safety = 0.9
\end{lstlisting}
where each integrator entry accepts \texttt{leapfrog}, \texttt{rk4}, or \texttt{rk45}. The RK45 parameters are used by any loop running in RK45 mode.

The time step and final time for the $\delta N$ calculation are specified separately:
\begin{lstlisting}[language=bash,numbers=none]
dN = 0.0000001
Nend = 0.001
\end{lstlisting}
Choosing these parameters requires some care. The step size \texttt{dN} must be small enough to resolve the evolution of the separate-universe equation (\cref{eq:eom_deltaN}). As a rule of thumb, it should be much smaller than the expected amplitude of the curvature perturbation, i.e. $\delta N \ll \sqrt{\mathcal{P}_{\zeta}}$. For very small $\mathcal{P}_{\zeta}$, the required step size becomes prohibitively small, and the $\delta N$ calculation is typically unnecessary—nonlinear effects are only important for $\mathcal{P}_{\zeta} \gtrsim 10^{-4}$, as shown in~\cite{Caravano:2025diq}.

The final time \texttt{Nend} should be long enough for all lattice points to reach a common final constant-field hypersurface $\phi_{\rm ref}$.
The stopping logic uses compile-time flags in \texttt{parameters.h}.
For monotonic regions, set \texttt{monotonic\_potential} to \texttt{1}.
For anti-monotonic regions, set \texttt{antimonotonic\_potential} to \texttt{1}.
If this behavior is clear, the code can use a faster field-based stopping test.

If the potential does not have a clear monotonicity in that region, set both flags to \texttt{0}: the code still handles the evolution correctly using the generic potential-based test, but this check is typically a bit slower.
This choice approximates a uniform-energy final slice when the dynamics is close to an attractor, but it may be problematic in non-attractor regimes or in models with multiple minima or extended plateaus.
In such cases, the user can manually define the hypersurface by uncommenting the following line:
\begin{lstlisting}[language=bash,numbers=none]
use_phiref_manual = 1
phiref_manual_value = 2.68
\end{lstlisting}
When enabled, the $\delta N$ calculation will use the manually specified reference field value instead of determining it dynamically.

\section{Scalar-induced gravitational waves}
\label{sec:GW}

\begin{figure}
    \centering
    \includegraphics[width=\linewidth]{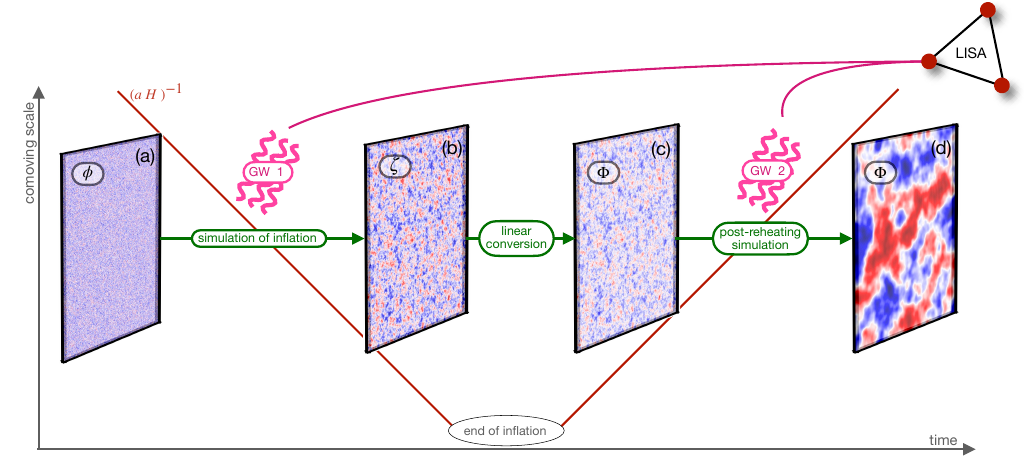}
    \caption{\textit{Scalar-induced gravitational-wave pipeline implemented in} \texttt{InflationEasy}.\\
    Starting from sub-horizon inflaton fluctuations evolved non-perturbatively on the lattice during inflation (a), the simulation yields the super-horizon curvature perturbation $\zeta$ at the end of inflation (b). This curvature field is then mapped to the Newtonian potential $\Phi$ on super-horizon scales using the initial relation given in \cref{eq:zeta_to_phi} (c), which provides the initial conditions for the post-inflationary evolution. The subsequent real-space evolution describes the horizon re-entry of scalar perturbations, yielding the sub-horizon Newtonian potential (d). The resulting gravitational-wave signal contains two contributions: an inflationary component, generated during inflation (GW $1$), and a post-inflationary component sourced at horizon re-entry (GW $2$).}
    \label{fig:gwpipeline}
\end{figure}

\texttt{InflationEasy} supports the calculation of scalar-induced gravitational waves (SIGWs), i.e.\ tensor perturbations generated at second order by scalar fluctuations~\cite{Tomita:1975kj,Matarrese:1993zf,Acquaviva:2002ud,Mollerach:2003nq,Ananda:2006af,Baumann:2007zm}. The code computes two physically distinct contributions to the resulting gravitational-wave (GW) background. One contribution is generated during inflation by the nonlinear dynamics of the inflaton field, while the other is generated after inflation, when the scalar perturbations produced during inflation re-enter the horizon in the post-inflationary phase. An overview of the computational pipeline and of the stages at which these contributions are generated is shown schematically in \cref{fig:gwpipeline}. While this section focuses on the numerical implementation in \texttt{InflationEasy}, the underlying physical framework and detailed discussion are presented in the companion physics paper~\cite{gwpaper}.

\subsection{Conventions and tensor equations}

\subsubsection*{Metric perturbations and normalization}

We consider tensor perturbations of a flat FLRW metric written as
\begin{equation}
\label{eq:GW_metric}
ds^2
=
a^2(\eta)\left[
- d\eta^2
+ \left(\delta_{ij} + h_{ij}(\eta,\vec{x})\right) dx^i dx^j
\right],
\end{equation}
where $\eta$ denotes conformal time and $h_{ij}$ is a symmetric tensor encoding the GW degrees of freedom. In this paper we use the convention above, without an additional normalization factor in front of $h_{ij}$. The companion physics paper~\cite{gwpaper} uses instead an extra $1/2$ factor in the definition of $h_{ij}$.

\subsubsection*{Tensor evolution equation}

Tensor perturbations are sourced at second order by scalar fluctuations. In the absence of anisotropic stress, the transverse--traceless (TT) tensor modes obey~\cite{Acquaviva:2002ud, Mollerach:2003nq, Ananda:2006af, Baumann:2007zm}:
\begin{equation}
\label{eq:GW_eom}
h_{ij}^{{\rm TT}\,\prime\prime}(\eta,\vec{x})
+2\mathcal{H}\,h_{ij}^{{\rm TT}\,\prime}(\eta,\vec{x})
-\nabla^2 h_{ij}^{\rm TT}(\eta,\vec{x})
=
2\,\mathcal{T}_{ij}^{\ \ \ell m}\,S_{\ell m}(\eta,\vec{x}),
\end{equation}
where $\mathcal{H}=a'/a$ is the conformal Hubble rate, primes denote derivatives with respect to conformal time, and $S_{ij}$ is the scalar-induced source term. The operator $\mathcal{T}_{ij}^{\ \ \ell m}$ denotes the TT projector acting on a symmetric tensor, and can be written as
\begin{equation}
\label{eq:GW_TT_operator}
\mathcal{T}_{ij}^{\ \ \ell m}
=
\frac12\left(
\mathcal{T}_i^{\ \ell}\mathcal{T}_j^{\ m}
+\mathcal{T}_j^{\ \ell}\mathcal{T}_i^{\ m}
-\mathcal{T}_{ij}\mathcal{T}^{\ell m}
\right),
\qquad
\mathcal{T}_{ij}
=
\delta_{ij}-\frac{\partial_i\partial_j}{\partial^2},
\end{equation}
with $\partial^2\equiv \partial_k\partial_k$. By construction, $h_{ij}^{\rm TT}$ satisfies $\partial_i h_{ij}^{\rm TT}=0$ and $h_{i}^{\ {\rm TT}i}=0$.

In Fourier space, the same projector is implemented mode by mode. Defining $\hat{k}=\vec{k}/|\vec{k}|$ and the transverse projector $P_{ij}(\hat{k})=\delta_{ij}-\hat{k}_i\hat{k}_j$, the TT projector reads
\begin{equation}
\label{eq:GW_TT_k}
\Lambda_{ij}^{\ \ \ell m}(\hat{k})
=
P_i^{\ \ell} P_j^{\ m}
-\frac12 P_{ij} P^{\ell m},
\end{equation}
so that for any symmetric tensor $X_{ij}$,
\begin{equation}
X_{ij}^{\rm TT}(\eta,\vec{k})
=
\Lambda_{ij}^{\ \ \ell m}(\hat{k})\,X_{\ell m}(\eta,\vec{k}).
\end{equation}

\subsection{SIGWs during inflation}

During inflation, the scalar-induced source term entering \cref{eq:GW_eom} is
\begin{equation}
\label{eq:GW_source_infl}
S_{ij}^{(\rm infl)}
=
\partial_i \phi(\eta,\vec{x})\,
\partial_j \phi(\eta,\vec{x}),
\end{equation}
where $\phi$ is the canonical inflaton field evolved on the lattice (see \cref{sec:eoms,sec:implementation}). This contribution captures GWs sourced during inflation by the nonlinear dynamics of the inflaton field.

\subsection{Post-inflationary SIGWs}

After inflation, \texttt{InflationEasy} can optionally perform a post-inflationary evolution in which scalar perturbations are described by the Newtonian potential $\Phi(\eta,\vec{x})$ in a perfect-fluid background with constant equation of state $w$. In this stage, $\Phi$ evolves according to the Bardeen equation
\begin{equation}
\label{eq:GW_Bardeen}
\Phi'' + 3\mathcal{H}(1+w)\Phi' - w\nabla^2 \Phi = 0.
\end{equation}
Initial conditions for the post-inflationary evolution are obtained from the super-horizon curvature perturbation $\zeta(\vec{x})$ produced by the inflationary lattice simulation. On super-horizon scales, $\zeta$ is conserved and is related to the Newtonian potential $\Phi$ through
\begin{equation}
\label{eq:zeta_to_phi}
\Phi(\eta \rightarrow 0,\vec{x})
=
\frac{3(1+w)}{5+3w}\,\zeta(\vec{x}),
\qquad
\Phi'(\eta \rightarrow 0,\vec{x}) = 0 ,
\end{equation}
which is used to initialize the post-inflationary evolution of $\Phi$ in the code.

The corresponding scalar-induced source term for tensor modes is
\begin{equation}
\label{eq:GW_source_post}
\begin{split}
S_{ij}^{(\rm post)}
&=
4\Phi\,\partial_i\partial_j\Phi
+2\,\partial_i\Phi\,\partial_j\Phi \\
&\quad
-\frac{4}{3(1+w)}
\partial_i\!\left(\frac{\Phi'}{\mathcal{H}}+\Phi\right)
\partial_j\!\left(\frac{\Phi'}{\mathcal{H}}+\Phi\right),
\end{split}
\end{equation}
which coincides with the standard second-order Newtonian-gauge source used in analytical and semi-analytical calculations of SIGWs. The post-inflation evolution is similar to the one used in~\cite{Ning:2026nfs}.\footnote{In Ref.~\cite{Ning:2026nfs}, the initial conditions are generated using an analytic local ansatz for the curvature perturbation, $\zeta = F(\zeta_g)$, rather than being obtained from a simulation of inflation.}

\paragraph{Assumptions and limitations.}
The post-inflationary calculation relies on a number of standard approximations. The background expansion is fixed and described by a perfect fluid with constant equation of state $w$, metric backreaction from tensor modes is neglected, and the scalar sector is evolved linearly at the level of the Newtonian potential $\Phi$. At the same time, in contrast to standard semi-analytical approaches, the initial conditions for the post-inflationary evolution are taken directly from the lattice simulation of inflation. As a result, all non-Gaussian features of the primordial curvature perturbation are retained to all orders and consistently propagated into the GW source. A detailed discussion of these assumptions, as well as of their validity and limitations, is given in the companion physics paper~\cite{gwpaper}.

\subsection{Tensor spectra and gravitational-wave energy density}
The tensor power spectra are defined through
\begin{align}
\label{eq:GW_Pdef}
\langle h_{ij}^{\rm TT}(\eta,\vec{k})\, h_{ij}^{\rm TT}(\eta,\vec{k}') \rangle
&= (2\pi)^3 \delta^{(3)}(\vec{k}+\vec{k}')\, 2P_h(\eta,k), \\
\langle h_{ij}^{{\rm TT}\,\prime}(\eta,\vec{k})\, h_{ij}^{{\rm TT}\,\prime}(\eta,\vec{k}') \rangle
&= (2\pi)^3 \delta^{(3)}(\vec{k}+\vec{k}')\, 2P_{h'}(\eta,k).
\end{align}
With this convention, $P_h$ and $P_{h'}$ are defined per tensor polarization (assuming parity symmetry between $+$ and $\times$), while $2P_h$ and $2P_{h'}$ correspond to the total spectrum summed over the two polarizations.
From these spectra we define the GW energy density per logarithmic interval as
\begin{equation}
\label{eq:GW_rho}
\frac{d\rho_{\rm GW}}{d\ln k}
=
\frac{1}{32\pi G\,a^2}
\frac{k^3}{2\pi^2}
\left[
P_{h'}(\eta,k) + k^2 P_h(\eta,k)
\right],
\end{equation}
and the corresponding fractional energy density as\footnote{This formula differs from the companion physics paper~\cite{gwpaper} because we use a different convention for the normalization of $h_{ij}$, as explained at the beginning of \cref{sec:GW}.}
\begin{equation}
\label{eq:GW_Omega}
\Omega_{\rm GW}(\eta,k)
=
\frac{1}{12\,a^2\mathcal{H}^2}
\frac{k^3}{2\pi^2}
\left[
P_{h'}(\eta,k) + k^2 P_h(\eta,k)
\right].
\end{equation}
These expressions are used throughout to characterize the GW signal generated both during inflation and in the post-inflationary phase. The quantity $\Omega_{\rm GW}$ admits its usual interpretation as a physical energy density once the corresponding modes are well inside the horizon, while the spectra $P_h$ and $P_{h'}$ are directly constructed from the lattice evolution at all times. The energy density spectrum $\Omega_{\rm GW}(k)$ is evaluated at the time of the calculation; to relate it to the present-day stochastic background, one typically assumes a standard thermal history. For example, if the GWs are emitted during radiation domination, as in the case studied in \cite{gwpaper}, the present-day value can be estimated as
\[
\Omega_{\rm GW,0}(k)\;\simeq\;\Omega_{\rm GW}(k)\,\Omega_{r,0},
\]
where $\Omega_{r,0}\simeq 9\times10^{-5}$ is the radiation density parameter today.

\subsection{Numerical implementation}
\label{sec:GW_numerics}

The GW sector is evolved on the same three-dimensional lattice, using the same time integrator as the scalar field at each stage (leapfrog, RK4, or RK45, depending on run-time settings). As in standard lattice implementations, the code evolves the six independent components of the symmetric tensor $h_{ij}(\eta,\vec{x})$ in real space, and applies the transverse--traceless (TT) projection only when constructing Fourier-space observables. This procedure is exact, because the TT projector commutes with the linear evolution in \cref{eq:GW_eom_code}~\cite{Garcia-Bellido:2007fiu}. Concretely, the six components $h_{ij}$ are evolved as:
\begin{equation}
\label{eq:GW_eom_code}
h_{ij}'' + 2\mathcal{H} h_{ij}' - \nabla^2 h_{ij}
=
2\,S_{ij},
\end{equation}
where $S_{ij}$ is the scalar-induced source term, given by \cref{eq:GW_source_infl} during inflation and by \cref{eq:GW_source_post} in the post-inflationary phase.

\subsubsection*{Fourier-space projection and effective momenta}

Spectra are computed by Fourier transforming each of the six components of $h_{ij}$ (and of $h_{ij}'$) and applying the TT projector in $k$-space. For each Fourier mode, the code constructs the transverse projector
\begin{equation}
P_{ij}(\hat{k})=\delta_{ij}-\hat{k}_i\hat{k}_j,
\qquad
\hat{k}=\vec{k}/|\vec{k}|,
\end{equation}
and the TT projector
\begin{equation}
\label{eq:GW_TT_k_num}
\Lambda_{ij}^{\ \ \ell m}(\hat{k})
=
P_i^{\ \ell} P_j^{\ m}
-\frac12 P_{ij} P^{\ell m}.
\end{equation}
Here $\vec{k}$ is not taken to be the naive lattice momentum $2\pi\vec{m}/L$, but the effective momentum associated with the second-order finite-difference Laplacian, consistent with the modified dispersion relation discussed in \cref{sec:modified}. Concretely, for each component the code uses the sine-based prescription (equivalent to \cref{eq:eff}) to build the components of $\vec{k}$ entering $\hat{k}$ and $\Lambda_{ij}^{\ \ \ell m}$. In this way, the same momentum convention is used for scalar spectra, for the TT projector, and for GW binning.

\subsubsection*{Mode binning and TT contraction}

The TT projection is implemented at the level of the power spectrum by contracting the Fourier-space tensor with $\Lambda$. For each Fourier mode, the code forms the complex symmetric matrix $h_{ij}(\eta,\vec{k})$ from the FFT output, computes the TT-projected tensor
\begin{equation}
h_{ij}^{\rm TT}(\eta,\vec{k})=\Lambda_{ij}^{\ \ \ell m}(\hat{k})\,h_{\ell m}(\eta,\vec{k}),
\end{equation}
and accumulates the squared norm $h_{ij}^{\rm TT}(\eta,\vec{k})\,h_{ij}^{\rm TT}(\eta,-\vec{k})$ into spherical bins in $|\vec{k}|$. As in the scalar sector, the bin centers are associated with the average effective momentum of the modes contributing to each bin (cf.\ \texttt{modes.dat} and the discussion in \cref{sec:output}). The code outputs GW spectra only up to 80\% of the Nyquist frequency to avoid bins dominated by ultraviolet lattice artifacts.

\subsubsection*{Time derivatives and code-unit conversion}

The code stores tensor time derivatives in the same rescaled time coordinate used for the scalar evolution (see \cref{eq:rescaling}). When constructing \texttt{spectraGWdot.dat}, the stored quantity is converted to the cosmic-time derivative $\dot h_{ij}$ in physical units. The TT projection and binning are then performed identically to the $h_{ij}$ case.

\subsection{Usage instructions}

\subsubsection*{Activating the GW calculation}

The computation of SIGWs is controlled by both compile-time and run-time parameters: feature activation is compile-time in \texttt{parameters.h}, while run-time settings (e.g.\ equation-of-state and integration controls) are set in \texttt{params.txt}. For the post-inflation stage, the step size, integrator, and final scale factor can be set with
\begin{lstlisting}[language=bash,numbers=none]
dt_post_inflation = ...
post_inflation_integrator = leapfrog
af_post_inflation = ...
\end{lstlisting}
If \texttt{af\_post\_inflation} is omitted, the default is
$
a_f^{\rm post} = 2N_{\rm pts}.
$

\subsubsection*{Outputs}

\begin{figure}
    \centering
    \includegraphics[width=0.49\linewidth]{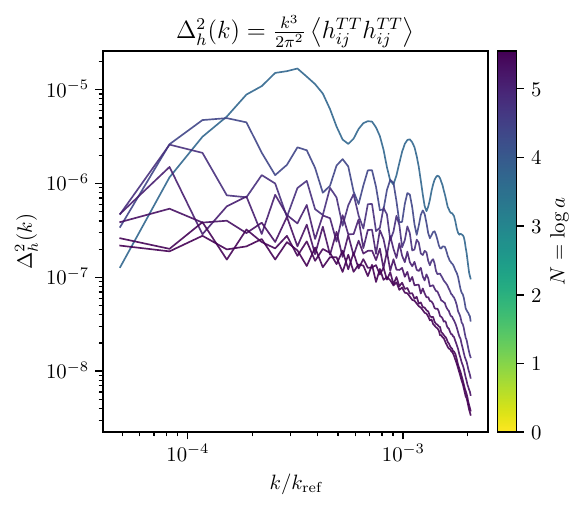}
    \includegraphics[width=0.49\linewidth]{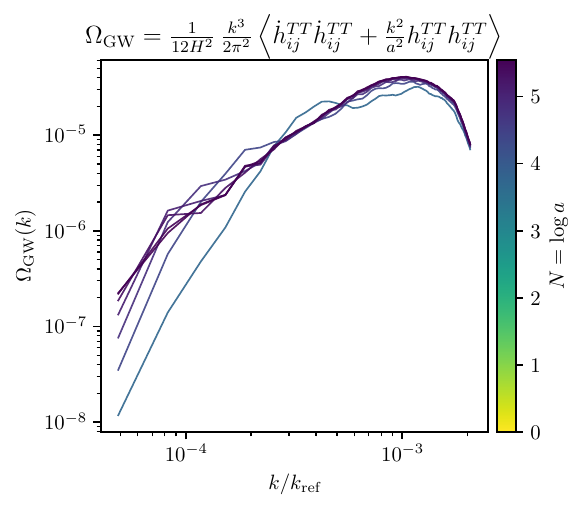}
    \caption{Post-inflationary dimensionless tensor power spectrum $\Delta_h^2(k)$ (left) and gravitational-wave energy density spectrum $\Omega_{\rm GW}(k)$ (right) in the default ultra-slow-roll (USR) example, obtained from the same simulation run used for \cref{fig}. The color scale indicates the number of e-folds $N=\log a$ from the beginning of the post-inflation simulation. These plots are generated using the example notebook \texttt{plot.ipynb} included in the public release.}

    \label{fig:gw}
\end{figure}

When enabled, the code outputs tensor power spectra constructed from the TT-projected tensor field and its time derivative. During inflation, the files
\begin{itemize}
    \item \texttt{spectraGW.dat},
    \item \texttt{spectraGWdot.dat}
\end{itemize}
are produced. If the post-inflationary evolution is enabled, analogous files are written in the directory
\begin{itemize}
    \item \path{post_inflation/spectraGW.dat},
    \item \path{post_inflation/spectraGWdot.dat}.
\end{itemize}
The \path{post_inflation/} directory also contains scalar-sector outputs.
These are associated with the post-inflationary evolution of the Newtonian potential $\Phi$.
The following files may be generated:
\begin{itemize}
    \item \textbf{Background files:}
    \path{post_inflation/sf.dat}, \path{post_inflation/means.dat},
    \path{post_inflation/velocity.dat}, and \path{post_inflation/variance.dat}.
    The file \path{post_inflation/sf.dat} has four columns: simulation time $\tilde{\tau}$, scale factor $a$, Hubble rate $H=\dot a/a$, and $\ddot a$. The files \path{post_inflation/means.dat}, \path{post_inflation/velocity.dat}, and \path{post_inflation/variance.dat} each have three columns (simulation time, scale factor, and the corresponding quantity), with the last column equal to $\langle \Phi \rangle$, $\langle \dot{\Phi} \rangle$, and $\langle \Phi^2 \rangle - \langle \Phi \rangle^2$, respectively.
    \item \textbf{Scalar spectra files:}
    \path{post_inflation/spectra.dat} and \path{post_inflation/spectratimes.dat}.
    These contain the power spectrum of the Newtonian potential $\Phi$, stored as a concatenated list of spectra evaluated at different times, and the corresponding evaluation times.
\end{itemize}
From these outputs, one can directly construct the dimensionless tensor power spectrum $\Delta_h^2(k)$ and the GW energy density spectrum $\Omega_{\rm GW}(k)$. In practice, with \texttt{spectraGWdot.dat} and \texttt{sf.dat}, the equivalent cosmic-time form is used:
\[
\Omega_{\rm GW}(k)=\frac{1}{12H^2}\frac{k^3}{2\pi^2}\left[P_{\dot h}(k)+\frac{k^2}{a^2}P_h(k)\right].
\]
Figure~\ref{fig:gw} shows a typical example of the resulting post-inflationary tensor spectra. An explicit example illustrating how to compute $\Omega_{\rm GW}$ from the raw output files and reproduce these plots is provided in the example notebook included with the public release.

\section{Conclusions and outlook}
\label{sec:conclusions}

\texttt{InflationEasy} provides a lightweight and accessible framework to simulate single-field cosmological inflation on a three-dimensional lattice. Unlike most existing lattice codes, which were primarily developed to study post-inflationary dynamics such as preheating, \texttt{InflationEasy} is tailored specifically to the inflationary phase itself. The code enables a fully nonlinear evolution of the inflaton field without relying on a background--perturbation split, making it well suited to investigate inflationary dynamics in regimes where nonlinear effects are important.

\texttt{InflationEasy} incorporates several features designed for inflationary applications. These include an improved initialization of Bunch--Davies fluctuations that accounts for the lattice-modified dispersion relation, a consistent treatment of Fourier-space observables, and the direct computation of the comoving curvature perturbation $\zeta$ via a fully nonlinear $\delta N$ calculation. Together, these ingredients allow for robust, nonperturbative calculations of inflationary observables in scenarios where nonlinear dynamics and large fluctuations play an essential role. The code has been tested across a range of models, including scenarios with enhanced small-scale fluctuations relevant for primordial black hole formation and gravitational wave backgrounds~\cite{Caravano:2024moy,Caravano:2024tlp,Caravano:2025diq,gwpaper}.

In addition to scalar observables, \texttt{InflationEasy} includes the computation of scalar-induced gravitational waves. The code accounts for contributions generated both during inflation and during the subsequent horizon re-entry of scalar perturbations, enabling fully nonlinear predictions of the resulting stochastic gravitational-wave background. This makes \texttt{InflationEasy} a unified numerical framework to study the joint scalar and tensor signatures of inflation within a single simulation setup.

A key strength of \texttt{InflationEasy} is its simplicity and accessibility. Despite being written in C++, the code is modular and easy to read, making it suitable even for users with limited programming experience. Compile-time options are configured in \texttt{parameters.h}, while most physical and numerical run settings are controlled through \texttt{params.txt}; inflationary potentials can be modified straightforwardly in \texttt{potential.cpp}.

Looking ahead, possible extensions of \texttt{InflationEasy} include support for multifield models and further numerical improvements, such as enhanced parallelization and higher-order spatial discretization schemes. We hope that \texttt{InflationEasy} will serve as a useful tool for both model builders and phenomenologists interested in the nonlinear dynamics of inflation and its observational signatures. We welcome feedback via the public repository at \href{https://github.com/caravangelo/inflation-easy.git}{https://github.com/caravangelo/inflation-easy.git}.

\section*{Generative AI Statement}

Generative AI tools were used to assist parts of the code documentation (comments and docstrings), to support implementation of some code features (including the RK45 integrator), and to assist language proofreading of the manuscript. All scientific analyses, numerical validation, methodological choices, and final code and manuscript content were reviewed and validated by the authors, who take full responsibility for this work.

	\section*{Acknowledgements}

A special thanks goes to S\'ebastien Renaux-Petel and the other organizers of the \href{https://indico.ijclab.in2p3.fr/event/11373/}{Cosmology Beyond the Analytic Lamppost} (CoBALt) workshop, held in June 2025 in Orsay, France. Preparing lectures for this workshop offered the perfect opportunity to initiate the public release of the code. The author also thanks Drew Jamieson, Eiichiro Komatsu, Kaloian D. Lozanov, Mauro Pieroni, Antonio Riotto and Denis Werth for insightful discussions. The author thanks Gabriele Franciolini, Kaloian D. Lozanov and S\'ebastien Renaux-Petel for comments on the draft.

This work was supported in part by the Initiative Physique des Infinis (IPI), a research training program of the Idex SUPER at Sorbonne Universit\'e. This project has also received funding from the European Union’s Horizon Europe research and innovation programme under the Marie Skłodowska-Curie grant agreement No. 101202657.

	\bibliographystyle{jhep}
	\bibliography{main}
	
\end{document}